\documentclass[aps,prl,floatfix,twocolumn]{revtex4}
\usepackage{dcolumn}
\usepackage{bm}
\usepackage[dvips]{epsfig}
\usepackage{graphicx} \usepackage{amsmath} \usepackage{amssymb}

\def\bea{\begin{eqnarray}}
\def\eea{\end{eqnarray}}
\def\dd{\partial}

\newcommand{\pa}{\partial}
\newcommand{\di}{\Delta^{ -1}}

\newcommand{\comment}[1]{}
\newcommand{\BEQ}{\begin{equation}}
\newcommand{\EEQ}{\end{equation}}
\newcommand{\BEA}{\begin{eqnarray}}
\newcommand{\EEA}{\end{eqnarray}}
\renewcommand{\d}{{\rm d}}

\renewcommand{\a}{\alpha}
\renewcommand{\b}{\beta}

\begin{document}

\title{Defining work done on electromagnetic field}

\author{A.E. Allahverdyan and D. Karakhanyan}
\affiliation{ Yerevan Physics Institute, Alikhanian Brothers Street 2, Yerevan 375036, Armenia}

\begin{abstract} The problem of defining work done on electromagnetic
field (EMF) via moving charges does not have a ready solution, because
the standard Hamiltonian of EMF does not predict gauge-invariant energy
changes. This limits applications of statistical mechanics to EMF.  We
obtained a new, explicitly gauge-invariant Hamiltonian for EMF that
depends only on physical observables. This Hamiltonian allows to define
thermodynamic work done on EMF and to formulate the second law for the
considered situation. It also leads to a direct link between this law
and the electrodynamic arrow of time, i.e. choosing retarded, and not
advanced solutions of wave-equations. Measuring the thermodynamic work
can give information on whether the photon mass is small but non-zero. 

\end{abstract}


\maketitle

\comment{

We would like to explain the importance of our results and their
relevance for PRL. It is well-known that Hamiltonian dynamics is crucial
for statistical mechanics. Hamiltonians not only govern time-evolution,
but they are also necessary for defining basic quantities of
thermodynamics, viz. work and heat. In this manuscript, we noted an
important caveat that arises in application of statistical mechanics to
electromagnetic field (EMF): the standard Hamiltonian of EMF is not a
gauge-invariant quantity, hence one cannot employ this Hamiltonian for
defining the work done by moving charges on EMF. This caveat does limit
the application of statistical mechanics to EMF. Since the caveat is not
well-known, we did explain it carefully in the manuscript. Researching
on this subject is especially relevant in the context of recent
expansion of interest in non-equilibrium statistical mechanics, which to
a large extent is based on notions of work and heat. 

Our main result is that we were able to obtain a new, explicitly
gauge-invariant Hamiltonian for EMF in the presence of charges. It
depends only on physical degrees of freedom and it agrees with the
well-known expression by Poynting, whenever the charges are absent, i.e.
for a free EMF. The found Hamiltonian allows to address several
fundamental issues. First of all, it allows to define the work done on
EMF and to come up with the formulation of the second law. We stress
that older (gauge-variant) definitions of the Hamiltonian did not lead
to such a formulation.  Second, it shows explicitly how the second law
relates to the electrodynamical arrow of time. Such a direct relation
was sought since the Ritz-Einstein debate, but it was so far elusive
because no reliable definition of work was obtained. Third, it appeared
that the definition of work is capable of distinguishing between the
proper (i.e. zero-mass) EMF and its massive version. This hints that the
experimental fact of the zero photon mass can have relations with
thermodynamics. 

In sum, we believe our results led to important relations between
electromagnetism and thermodynamics, and we opine that they deserve to
be considered in PRL.

}

{\it Introduction.} Hamiltonian dynamics is essential for statistical
mechanics and thermodynamics \cite{gi}. Basic distribution functions of
statistical mechanics (e.g. canonical or microcanonical) are formulated
in the phase-space and are based on the conservation of energy and of
the phase-space volume (the Liouville's theorem)
\cite{gi,balian,lindblad,mahler}. Also the basic quantities of
thermodynamics|energy, work and heat|are defined via the Hamiltonian of
the system; e.g. the change of the time-dependent Hamiltonian for a
thermally isolated system defines the work done externally
\cite{balian,lindblad,mahler}. The first law divides energy into work
and heat \cite{balian}, while the second law limits work-extraction via
cyclic processes \cite{lindblad}. The third law studies work as a
resource for cooling \cite{cooling}. 

Our aim is to understand thermodynamic work done by moving charges
(sources of work) on electromagnetic field (EMF); the research done
on EMF from various angles (field-theoretic, quantum, statistical {\it
etc}) is reflected e.g. in \cite{landau,ribaric,photon,mass,stue}
\cite{comment}. We stress that thermodynamics and electrodynamics share
at least two structural features. {\it (1.)} Both study systems with
many degrees of freedom. {\it (2.)} Both need specific subsystems
(work-sources) whose motion is prescribed in the sense that the
back-reaction on them is partially neglected \cite{comment2}. For
thermodynamics these are e.g. vessels of a gas \cite{balian}, while for
EMF these are moving charges \cite{landau}. 

Given these similarities, the work done on EMF by moving charges is to
be defined via the Hamiltonian of EMF. For stationary charges the
Hamiltonian is conserved; hence there is no thermodynamic work.  Now the
standard Hamiltonian for EMF is conserved for stationary motion of
charges \cite{dirac,gitman}. But it appears that (for non-stationary
charges) the {\it change} of this Hamiltonian of EMF is not
gauge-invariant. Hence we cannot apply it for defining work.  After
discussing this issue, we determine an explicitly gauge-invariant
Hamiltonian of EMF that {\it (i)} generates Maxwell's equations via
Hamiltonian equations, {\it (ii)} reduces to the standard expression for
the free EMF, {\it (iii)} allows to define thermodynamic work done on
EMF. This work consists of electrostatic and vortical contributions.
{\it (iv)} If the work done on EMF is measured independently, it can
indicate on whether the mass $m$ of photon holds $m=0$ or $m>0$.  {\it
(v)} The definition of work demonstrates an explicit relation between
thermodynamic arrow of time (i.e. the second law) and the electrodynamic
arrow of time. Despite of opinions expressed since the Ritz-Einstein
debate \cite{debate}, the two arrows are so far regarded to be different
from each other \cite{cramer,fritz,zeh_book,frisch}.

{\it The Lagrangian of a classical EMF} for a given motion of charged matter 
with density $\rho$ and current $J_i$ reads \cite{landau}
\begin{align}
\label{lagro}
&L=\int{\rm d}^3\,x\, {\cal L},\quad {\cal L}[\phi, A_i]=\frac{E_i^2}{2}-\frac{B_i^2}{2}-\rho\phi+J_iA_i,\\
\label{electro}
&E_i=-\partial_i \phi-\dot A_i, \quad i=1,2,3, \\
\label{magno}
&B_i=\epsilon_{ijk}\partial_j A_k,~ \epsilon_{ikl} B_l=\pa_i A_k-\pa_k A_i, 
\end{align}
where $E_i$ and $B_i$ are (resp.) electric and magnetic fields, $\phi$
and $A_i$ are (resp.) scalar and vector potential. We took $c=1$, 
denoted the 3d coordinate as $x=(x_1,x_2,x_3)$ (e.g. $E_i=E_i(x,t)$).
Repeated space-indices imply summation, 
$\partial_i\equiv{\partial}/{\partial x_i}$, $\dot A_i\equiv\partial_t
A_i$, and $\epsilon_{ijk}$ is
the totally asymmetric factor with $\epsilon_{123}=1$.

Now $L$ refers to coordinates $\phi(x,t)$ and $A_i(x,t)$ and velocities
$\dot A_i(x,t)$ that are parametrized by a continuous index $x$ and discrete
index $i$. Hence the Lagrange equations deduced from $L$ have the usual
form, but with variational derivatives
\BEA
\frac{\d}{\d t}[\,{\delta L }\left/{\delta \dot A_j(y) }\right.\,]={\delta L }\left/{\delta A_j(y) }\right.,\quad
{\delta L }\left/{\delta \phi(y) }\right.=0.
\label{mak}
\EEA
Note that ${\cal L}$ does not contain $\dot\phi$; hence the last equation in (\ref{mak}).
When working out (\ref{mak}) we standardly assume that $\rho$, $J_i$,
$E_i$ and $B_i$ decay to zero at the spatial infinity, apply integration
by parts, and employ known formulas of variational calculus, e.g.
${\delta A_i(x) }\left/{\delta A_j(y) }\right.=\delta_{ij}\,\delta(x-y)$
with Kronecker and Dirac's deltas, respectively. Hence we get from
(\ref{mak}) equations of motion:
\bea
\label{agnus1}
&&\partial_k \dot\phi  +\ddot A_k=\Delta A_k-\pa_k (\pa_i A_i)+J_k,\\
&&\Delta\phi=-\rho-\pa_i\dot A_i,
\label{agnus2}
\eea
where $\Delta=\pa_i\pa_i$ is the Laplace operator. Eqs.~(\ref{electro}, \ref{magno}) show that 
(\ref{agnus1}) and (\ref{agnus2}) become (resp.) the Maxwell's equations
\bea
\label{max}
\dot{E}_i=\epsilon_{ijk}\pa_jB_k-J_i,\qquad
\pa_i E_i=\rho.
\eea
Eqs.~(\ref{agnus1}, \ref{agnus2}) also imply the conservation of charge:
\BEA
\label{cons}
\dot\rho+\pa_k J_k=0.
\EEA

{\it The standard Hamiltonian of EMF} is constructed from (\ref{lagro}).
One should note here that strictly speaking the EMF is a singular
system, since ${\cal L}$ does not contain $\dot \phi$ \cite{dirac,gitman}. This
singularity can be dealt with in various equivalent ways, also via the
full Dirac's formalism \cite{dirac,gitman}. But the simplest way
is to carry out the Legendre transformation with respect to $\dot
A_i$ only \cite{dirac,gitman}:
\bea
\label{ven}
H_D=\int{\rm d}^3\,x\,{\cal H}_D,\quad 
{\cal H}_D=p_i\dot{A}_i - {\cal L},
\eea
where the canonic momentum $p_i$ is defined from
\bea
{\delta H_D}\left/{\delta \dot A_k(y)}\right.=0 \quad {\rm or}\quad
\dot A_i=p_i-\pa_i\phi.
\label{gamel}
\eea
Putting (\ref{gamel}) into ${\cal H}_D$, and making integration by parts we arrive at 
\cite{dirac,gitman}:
\bea
\label{kagan}
{\cal H}_D=\frac{1}{2}p_i^2+\frac{1}{2}B_i^2-J_iA_i +\phi(\pa_ip_i+\rho),
\eea
where $\phi$ is now the Lagrange multiplier for the constraint
$\pa_ip_i+\rho=0$ (given also by (\ref{mak}, \ref{agnus2}, \ref{gamel})).
Hamilton equations of motion are read from (\ref{kagan}) 
with canonic coordinates $A_i$, momenta $p_i$
and the Lagrange factor $\phi$ \cite{dirac,gitman}:
\BEA
\dot{A}_i={\delta H_D}/{\delta p_i}, \quad
\dot{p}_i=-{\delta H_D}/{\delta A_i}, \quad
{\delta H_D}/{\delta \phi}=0.
\label{hamo}
\EEA
Eqs.~(\ref{hamo}) bring back (\ref{agnus1}, \ref{agnus2}). On solutions
of (\ref{agnus1}, \ref{agnus2}), we have $E_i=-p_i$ from (\ref{electro},
\ref{gamel}). Hence ${\cal
H}_D=\frac{1}{2}E_i^2+\frac{1}{2}B_i^2-J_iA_i$, which for $J_i=0$
reduces to the well-known \cite{landau}
\BEA
{ H}_{\rm free}
=\frac{1}{2}\int{\rm d}^3\,x\,[\,E_i^2+B_i^2\,].
\label{larmor}
\EEA
Now $H_D$ is generally time-dependent due to $\rho$ and $J_i$.
As for any time-dependent Hamiltonian, we have 
\BEA
\label{thou}
\dot H_{D}&=&\int{\rm d}^3\,x\,\left[\dot A_i\,\frac{\delta H_D}{\delta A_i}
+\dot p_i\,\frac{\delta H_D}{\delta p_i}+\dot \phi\,\frac{\delta H_D}{\delta \phi} 
\right]\\
&+&\int{\rm d}^3\,x\,\left[\phi\dot\rho-\dot J_iA_i\right].
\label{ser}
\EEA
Now (\ref{thou}) nullifies due to (\ref{hamo}), so $\dot H_{D}$
is determined by (\ref{ser}). Hence $H_D$ is conserved
if $\dot\rho=\dot J_i=0$, where the Lagrangian (\ref{lagro}) 
is time-translation invariant. Eq.~(\ref{ser}) could be guessed directly from (\ref{lagro}).

But we cannot apply $H_D$ and (\ref{ser}) for calculating energy change.
Recall that equations of motion (\ref{agnus1}, \ref{agnus2}, \ref{max}) are
invariant with respect to gauge change
\bea
\label{gaug}
\phi\to\phi+\dot\chi, \qquad A_k\to A_k-\pa_k\chi, 
\eea
where $\chi(x, t)$ is arbitrary. This invariance relates to the zero
mass of EMF \cite{mass}. Due to (\ref{cons}), the Lagrangian
(\ref{lagro}) changes under (\ref{gaug}) by a full time-derivative:
$L\to L-\frac{\d}{\d t}\int{\rm d}^3\,x\,\rho\chi$. 
Eq.~(\ref{ser}) also changes by a full time-derivate
under the gauge-change (\ref{gaug}) 
\bea
\dot H_{D}\to\dot H_{D}
+\frac{{\rm d}}{{\rm d} t}\int{\rm d}^3\,x\,\dot\rho\chi,
\label{bu}
\eea
where we used (\ref{cons}). For a Lagrangian a shift by a full
time-derivatives is allowed \cite{lagron}, but for a Hamiltonian it is a
problem, since it alters the energy change $\int_{t_1}^{t_2}\dot H_{D}\d
t$ between $t_1$ and $t_2$. Now $\dot H_{D}$ is gauge-invariant for a
particular case $\dot\rho(x,t_1)=\dot\rho(x,t_2)=0$ for all $x$. This is
too restrictive for the definition of the energy change and work.
Indeed, in a standard task of thermodynamics a many-body system (e.g.
EMF) is employed as an energy storage, i.e. the time-dependent
parameters are driven by different sources that exchange work through
the system. For such cases it is simply necessary to calculate the
energy change up to a given time, because this is the work that goes to
one of the work-sources. 

The gauge-variant $\dot H_D$ is not suitable for defining
work. 

{\it Gauge-invariant Hamiltonian.} We now assume that (together with
$E_i$ and $B_i$) also $\phi$ and $A_i$ decay to zero for $|x|\to\infty$.
This assumption imples a partial gauge-fixing [cf.~(\ref{gaug})], but
our final results will not depend on it.  Now (\ref{agnus2}) is solved
via the inverse Laplacian $\di$ as
\bea
\label{aba1}
&& \phi=-\di(\rho+\pa_i\dot A_i), \\
&& \di f(x)\equiv -\frac{1}{4\pi}\int {\rm d}^3\,y\, \frac{f(y)}{|x-y|},
\label{aba2}
\eea
where we note that the solution of the homogeneous Laplace equation
(that could appear in the RHS of (\ref{aba2})) nullifies due to 
assumed boundary conditions: $\phi(x)\to 0$ for $|x|\to\infty$. 
Note in (\ref{aba1}) that for $\di(\pa_i\dot A_i)$ to be finite 
it is necessary that $\dot A_i$ decays to zero at infinity, 
which we already assumed. 

\comment{ Here are two transformation patterns
\begin{align}
&-\rho\phi=\rho\di(\rho+\pa_i\dot A_i)\mapsto\rho\di\rho+(\pa_i J_i)\,\di (\pa_k A_k)\nonumber\\
\label{sense}
&\mapsto \rho\di\rho-J_iA_i - J_i\,\di (\pa_k[\pa_i A_k-\pa_k A_i]),\\
&\pa_i\phi+\dot{A}_i  
= -\di(\pa_i\rho+\epsilon_{ikl}\pa_k \dot{B}_l).
\label{karam}
\end{align}
where the first $\mapsto$ means that we neglected the
full time-derivative, while the second $\mapsto$ means that we applied
integration by parts.}

We put back (\ref{aba1}) into ${\cal L}$ trying to find a Lagrangian for
$A_i$. In subsequent calculations we shall employ (\ref{cons}),
\ref{aba1}, \ref{aba2}),$\epsilon_{ijk}
\epsilon_{inm}=\delta_{nj}\delta_{km}-\delta_{nk}\delta_{mj}$,
and a commutativity relation $\dd_k\di=\di\dd_k$, which holds when
acting on functions decaying at infinity; e.g. 
\begin{align}
\pa_i\phi+\dot{A}_i = -\di(\pa_i\rho+\epsilon_{ikl}\pa_k \dot{B}_l).
\label{karam}
\end{align}
We also neglect one full time-derivative (allowed for a Lagrangian), and also
full space-derivatives, due to assumed boundary conditions. 
After some transformations, see section 1 of \cite{sup}, we get a Lagrangian that instead of $A_i$
depends directly on the magnetic field $B_i$:
\begin{align}
\label{alef}
&L_B=\int{\rm d}^3\,x\,{\cal L}_B,\\
&{\cal L}_B=\frac{1}{2}{\rho\di\rho} -\frac{1}{2}{\dot B_i\di(\dot B_i)}-\frac{1}{2}{B_iB_i} -B_i\di(R_i),\nonumber\\
&R_i\equiv \epsilon_{ijk}\pa_jJ_k.
\label{rotor}
\end{align}
Eq.~(\ref{alef}) comes with a constraint that follows from (\ref{magno})
\bea
\pa_j B_j=0,
\label{betan}
\eea
and confirms that EMF has two independent coordinates.
 
In equations of motion $\frac{\d}{\d t}\frac{\delta L_B}
{\delta \dot B_k(y)}=\frac{\delta L_B}{\delta B_k(y)}$ we use
\bea
\label{gamet}
{\delta L_B}\left/{\delta \dot B_k(y)}\right.=-\di(\dot B_k)(y).
\eea
This leads to an autonomous equations for 
${B}_i$ that can be also
derived from the Maxwell's equations (\ref{max})
\bea
\ddot{B}_i - \Delta {B}_i -R_i=0.
\label{tuk}
\eea
Using (\ref{gamet}) we introduce the canonical momentum
$\Pi_k(x)=-\di(\dot B_k)(x)$,
and construct from (\ref{alef}) the Hamiltonian via the usual Legendre transformation
\begin{align}
&H_B=\int{\rm d}^3\,x\,{\cal H}_B(x)= \int{\rm d}^3\,x\,\left[\Pi_k\dot B_k - {\cal L}_B \right],\\
&{\cal H}_B=
-\frac{\rho\di\rho}{2} -\frac{\Pi_i\Delta\Pi_i}{2} +\frac{B_iB_i}{2}+B_i\di(R_i),
\label{golem}
\end{align}
where constraint (\ref{betan}) is implied. 


{\it (i)} Eq.~(\ref{tuk}) can be reproduced from (\ref{golem}) via
Hamilton equations $\dot\Pi_k=\frac{\delta H_B}{\delta B_k}$ and $\dot
B_k=-\frac{\delta H_B}{\delta \Pi_k}$. 

{\it (ii)} Though (\ref{golem}) depends only on the magnetic field $B_i$
and its derivatives, it is consistent with (\ref{larmor}): apply
$\epsilon_{nmi}\pa_m$ to both sides of the Maxwell's equation $\epsilon_{ijk}\pa_jE_k=-\dot
B_i$ (deduced from (\ref{electro}, \ref{magno})) and employ there
(\ref{max}). Then we can express $E_i$ via $\dot B_k$ and $\pa_i\rho$:
\BEA
\label{eavto}
E_i=\di(\pa_i\rho+\epsilon_{ijk}\pa_j\dot B_k).
\EEA
We put (\ref{eavto}) into (\ref{larmor}) and integrate by parts:
\BEA
{ H}_{\rm free}=\frac{1}{2}  \int{\rm d}^3\,x\,[
-\rho\di\rho -\dot B_i\di(\dot B_i)
+{B_i^2}\,],
\label{larmori}
\EEA
i.e. (\ref{golem}) for $R_i=0$ agrees with (\ref{larmor}). 
In particular, (\ref{larmori}) includes 
the case of free (and generically space-localized) EMF fields.

{\it (iii)} We define work done by charges via the standard
formula accepted in statistical mechanics \cite{balian}:
\BEA
\dot H_B=\int{\rm d}^3\,x\,[-\frac{1}{2}
\frac{{\rm d}}{{\rm d} t}(\rho\di\rho)+B_i\di(\dot R_i)],
\label{dan}
\EEA
where we employed the same method as in (\ref{thou}). Now $\dot H_B$
consists of two parts: the electrostatic due to $\rho\di\rho$ (see
section 2 of \cite{sup}) and vortical due to $\di R_i$;
cf.~(\ref{rotor}).  We stress that the electrostatic contribution does
not depend on fields, it depends only on the externally controlled
$\rho(x,t)$. But we kept it, e.g. because it allows (\ref{golem}) to agree with
a well-accepted expression (\ref{larmor}). Section 4 of \cite{sup}
shows that $H_B$ is conserved if $\dot J_k=0$ and $\rho$ is demanded to
be bounded for all times. 

To avoid confusions note that the work done on EMF according to
(\ref{dan}) does not directly relate to radiation, e.g. (\ref{dan}) is
zero for a rectilinear motion of charges [cf.~(\ref{rotor})], where we do
expect radiation if this motion is accelerated \cite{landau}. Indeed, in
the considered set-up, where fields nullify at infinity, the radiation
is always a part of fields, the one that has a specific asymptotics far
from charges \cite{landau}. On the other hand, if (\ref{dan}) is
non-zero, then there is certainly acceleration and hence radiation. 
Note that we always deal with the full energy
(space-integrated) energy of EMF. The localization of this energy is not
studied; this is another (and more difficult) problem \cite{armen}. 

{\it (iv)} What if photon has a small but non-zero mass $m$? Due to its
foundational importance, this question ponders in physics for decades
\cite{mass,stue}.  Experiments put stringent bounds on $m$ \cite{mass},
but they cannot show that $m=0$.  Even within such bounds $m>0$ can be
relevant e.g. in cosmology \cite{cosmo}. We show that $m>0$ leads to a
different definition of work. Recall that massive electrodynamics is a
consistent theory \cite{mass,stue} (see section 3.1 of \cite{sup}) that
amounts to adding to ${\cal L}$ in (\ref{lagro}) the massive term
$\frac{m^2}{2}(\phi^2-A_i^2)$. This changes equations of motion
(\ref{agnus1}, \ref{agnus2}) by adding $-m^2A_k$ to the RHS of
(\ref{agnus1}) and $m^2\phi$ to the RHS of (\ref{agnus2}). New equations
produce $\dot\rho +\pa_kJ_k=m^2(\dot\phi+\pa_kA_k)$.  Hence the
charge-conservation (\ref{cons}) {\it and} $m>0$ lead to the Lorenz
gauge $\dot\phi+\pa_kA_k=0$ \cite{mass}. Then (\ref{ser}) still applies
for the change of the total Hamiltonian, but now no gauge-change
(\ref{gaug}) can be made. Hence (\ref{ser}) is consistent for $m>0$.
Moreover, for $m>0$ the method of (\ref{aba1}--\ref{karam}) can be
generalized, but it does not lead to a Lagrangian (or Hamiltonian)
description of $A_i$; see section 3.2 of \cite{sup}. Hence (\ref{ser})
and (\ref{dan}) provide consistent {\it and different} definitions of
work for (resp.) $m>0$ and $m=0$. If the work done on EMF can be
measured independently, this will show whether or not the photon has a
mass. 

\comment{We stress that a general solution of (\ref{tuk}) is a linear
combination of the retarded and advanced solutions \cite{landau}; we
select the retarded solution as a manifestation of the electrodynamic
arrow of time \cite{zeh_book}. }

{\it Arrows of time.} We apply (\ref{dan}) to the second law.  We assume
that $R_i$ is switched on at some initial time, {\it and} there were no
free fields before that time: $R_i({x}, t)=B_i(x,t)=0$ for $t\leq 0$.
Given these initial conditions, (\ref{tuk}) shows that $B_i(x,t)$ for
$t>0$ can be related to $R_i$ via the {\it retarded} solution
\cite{landau}
\bea
\label{reto}
B_i({x}, t)=\frac{1}{4\pi}\int\frac{{\rm d}^3\,y}{|{x}-{y}|} \, 
R_i({y}, t-|{x}-{y}|).
\eea
We get from (\ref{dan}, \ref{reto})
\bea
&&\dot W\equiv \int{\rm d}^3\,x\, B_i\di(\dot R_i)({x}, t)
=-\frac{1}{(4\pi)^2}\int{\rm d}^3\,x\,\times \nonumber\\
&&\int\frac{{\rm d}^3\,y \,R_i({y}, t-|{x}-{y}|) }{|{x}-{y}|} 
\int\frac{{\rm d}^3\,z \,\dot R_i({z}, t) }{|{x}-{z}|}. 
\label{tudor}
\eea
We calculate (\ref{tudor}) in the non-relativistic limit \cite{landau};
cf. section 5.2 of \cite{sup}.  It assumes that $R_i({{y}, t})$ as a
function of ${y}$ is well-localized in the vicinity of (say) ${y}=0$;
e.g. $R_i({{y}, t})\simeq f_i(t)\delta(y)$.  Using this in the RHS of
(\ref{tudor}), and going to spherical coordinates in $\int{\rm
d}^3\,x\,$, we end up with
\bea
\label{lester}
\dot W= 
-\int_0^\infty\frac{ {\rm d}\,r}{4\pi}
\int{\rm d}^3\,y \,R_i({y}, t-r) 
\int{\rm d}^3\,z \,\dot R_i({z}, t).
\eea
Recall that all fields vanish at infinity
and that $R_i({y}, t)=0$ for $t\leq 0$.
We get from (\ref{lester}):
\bea
\dot W= 
-\frac{\chi_i(t)\ddot\chi_i(t)}{4\pi},~~ 
\chi_i(t)\equiv\int{\rm d}^3\,x \,\int_0^t {\rm d}\,s\,R_i({x},s),
\label{lankaster}
\eea
Now (\ref{dan}, \ref{lankaster}) imply for the energy change:
\bea
\left\{H_B\right\}_0^t =\left\{E_S\right\}_0^t +
\int_0^t \frac{{\rm d}\,s}{4\pi}\, [\,\dot\chi_i(s)]^2 - \frac{1}{4\pi} \chi_i(t)\dot\chi_i(t),
\label{hugo}
\eea
where $E_S=-\frac{1}{2}\int{\rm d}^3\,x\,\rho\di\rho$ is the electrostatic
energy and $\{X \}_0^t\equiv X(t)-X(0)$. Eq.~(\ref{hugo}) makes a
thermodynamic sense: $\left\{H_B\right\}_0^t$ is the work, which
for the considered thermally isolated system (EMF) is defined via its
Hamiltonian. $\left\{E_S\right\}_0^t$ is the part of energy that depends
only on the state of the system at times $0$ and $t$, but does not
depend on the trajectory. Hence it accounts for the reversible work. If
we impose the cyclicity condition, assuming that besides $R_i({y}, t)=0$
for $t\leq 0$, it also holds $R_i({y}, \tau)=0$, then the last terms in
(\ref{hugo}) vanishes at $t=\tau$ due to $\dot\chi_i(\tau)=0$. Hence we
get the statement of the second law: the irreversible work
$\left\{H_B-E_S\right\}_0^\tau=\{W\}_0^\tau $ is non-negative, i.e. the energy is put
into EMF. Here the validity of this statement relates to the definition
(\ref{dan}) and our assumption on localized $R_i(x,t)$. We stress that
such a statement is not be deduced from (\ref{ser}), even if
we assume one of standard gauges (e.g. the Lorenz gauge); 
see section 5 of \cite{sup}. 

Above derivation was done assuming initial conditions. Alternatively, we 
can employ final conditions assuming that
$R_i(x,t)=B_i(x,t)=0$ for $t>\tau$. Then the connection between
$R_i(x,t)$ and $B_i(x,t)=0$ for $t<\tau$ is to be given via the advanced
solution of (\ref{tuk}):
\bea
\label{advo}
B^{\rm [ad]}_i({x}, t)=\frac{1}{4\pi}\int\frac{{\rm d}^3\,y}{|{x}-{y}|} \, 
R_i({y}, t+|{x}-{y}|).
\eea
The fact that normally one employs retarded solution (\ref{reto}) 
via initial conditions, and not the advanced solution (\ref{advo}) via
final conditions amounts to the electrodynamic arrow of time
\cite{cramer,fritz,zeh_book,frisch}.  

Repeating the above steps and imposing the cyclicity condition 
$R_i(x,t)=0$ for $t<0$, we get [cf.~(\ref{lankaster})]
\bea
\left\{H_B-E_S\right\}_0^\tau =-\int_0^\tau \frac{{\rm d}\,s}{4\pi}\, 
\left[\,\dot\chi_i(s)\right]^2.
\label{victor}
\eea
Now instead of the second law we got its opposite: the energy is
extracted from EMF.  This links the thermodynamic arrow of time (second
law or putting work into the many-body system) and the electrodynamic
arrow. Relations between the cosmological and thermodynamical arrows 
were recently explored in \cite{dark}.

{\it In sum}, we found a new gauge-invariant Hamiltonian for
electromagnetic field (EMF) that holds all desiderata for defining work.
In particular, it leads to the second law (in contrast to other
definitions), relates it with the electrodynamic arrow of time, and
differs from the Hamiltonian obtained in the limit of vanishing photon
mass. Elsewhere, we shall quantize this Hamiltonian and explore its
consequences for quantum electrodynamics. 

\acknowledgements

We thank V.G. Gurzadyan and S.G. Babajanyan for discussions. The work
was supported by SCS of Armenia, grants 18RF-002 and 18RF-015.


\section{Supplementary Material}

\subsection{1. Derivation of $L_B$}

Let us recall the Lagrangian of EMF with sources
\begin{gather}
{\cal L}=\frac{1}{2}(\partial_i \phi+\dot A_i)^2-\frac{1}{2}B_i^2
-\rho\phi+J_iA_i,
\label{lagro2}
\end{gather}
as well as the relation between $\phi$ and $A_i$ via the inverse Laplacian
[cf. (1--3) and (18, 19) of the main text]:
\bea
\label{babo2}
&& \phi=-\di(\rho+\pa_i \dot A_i), \\ 
&& (\di f)({x})=-\int\frac{{\rm d}^3 y}{4\pi} \,\frac{f(y)}{|{x}-{y}|}.
\eea
We recall that derivatives commute with the inverse Lagrangian:
\BEA
\label{lolo}
\frac{\pa}{\pa x_i}(\di f)({x})
=-\int\frac{{\rm d}^3 y}{4\pi} \,\frac{1}{|{x}-{y}|} \, \frac{\pa f(y)}{\pa y_i},
\EEA
where it is assumed that $f(x)\to 0$ for $|x|\to 0$.
We shall write (\ref{lolo}) as 
\BEA
\pa_i\di=\di\pa_i, 
\EEA
and employ it freely.

We integrate (\ref{lagro2}) by parts and write it as
\bea
\label{lagro3}
\int\d^3x\,{\cal L}=\int\d^3x\,[\, -\frac{1}{2}\phi\Delta\phi -\rho\phi -\phi\pa_i\dot A_i \\
+\frac{1}{2}\dot A_i^2 -\frac{1}{2}B_i^2+J_iA_i\,].
\eea
Now the first three terms in the RHS of (\ref{lagro3}) are to be transformed via (\ref{babo2}),
and 
\bea
\label{oli2}
\int{\rm d}^3 x\,a(x) (\di b)(x) = \int{\rm d}^3 x\,b(x) (\di a)(x).
\eea
We get
\begin{gather}
\label{mmm2}
\int\d^3x\,{\cal L}_=\int\d^3x\,[\,
\frac{1}{2}(\pa_i\dot A_i)\Delta^{-1} (\pa_k\dot A_k)+ (\pa_i\dot A_i)\Delta^{-1}(\rho) \nonumber\\
-\frac{1}{2}\rho\Gamma^{-1}\rho +\frac{1}{2}\dot A_i^2 -\frac{1}{2}B_i^2+J_iA_i\,].
\end{gather}
Now we transform $(\pa_i\dot A_i)\Delta^{-1}(\rho)$. We write it as
\BEA
(\pa_i\dot A_i)\Delta^{-1}(\rho)=\frac{\d}{\d t}[\,(\pa_iA_i)\Delta^{-1}(\rho)\,]-(\pa_i A_i)\Delta^{-1}(\dot\rho),
\EEA
neglect the full time-derivative, employ the charge conservation $\dot \rho+\pa_k J_k=0$
and replace $(\pa_i\dot A_i)\Delta^{-1}(\rho)$ by $(\pa_i A_i)\Delta^{-1}(\pa_k J_k)$. Hence the transformed Lagrangian reads:
\begin{gather}
\int\d^3x\,{\cal L}_B =\int\d^3x\,[\,-\frac{1}{2}\rho\Delta^{-1}\rho +\frac{1}{2}(\pa_i\dot A_i)\Delta^{-1} (\pa_k\dot A_k) \nonumber\\
+ (\pa_i A_i)\Delta^{-1}(\pa_k J_k)+\frac{1}{2}\dot A_i^2 
-\frac{1}{2}B_i^2+J_iA_i].
\label{bingo2}
\end{gather}
Now recall definitions of the magnetic field $B_i$ and of the vorticity of the charge flow $R_i$:
\BEA
\label{bra}
&B_i=\epsilon_{ijk}\partial_j A_k,~ \epsilon_{ikl} B_l=\pa_i A_k-\pa_k A_i, \\
&R_i\equiv \epsilon_{ijk}\pa_jJ_k.
\label{tok}
\EEA
The following relations are deduced via integration by parts, (\ref{lolo}) and (\ref{bra}, \ref{tok}):
\begin{gather}
-\frac{1}{2}\int\d^3x\,\dot B_i \Delta^{-1}\dot B_i \nonumber\\
\label{kot2}
=\frac{1}{2}\int\d^3x\,[\, (\pa_k\dot A_k) \Delta^{-1}(\pa_i\dot A_i)+\dot A_k^2 \,],\\
\int\d^3x\,(\pa_i A_i)\Delta^{-1}(\pa_k J_k) \nonumber\\
=-\int\d^3x\,A_i\Delta^{-1}(\pa_k [\pa_i J_k-\pa_k J_i+ \pa_k J_i])\nonumber\\
=-\int\d^3x\,[A_iJ_i+B_i\di R_i].
\label{hot2}
\end{gather}
Putting (\ref{kot2}, \ref{hot2}) into (\ref{bingo2}) we see that 
$\int\d^3x\,{\cal L}_B$ can be expressed only via $B_i$ and $R_i$. 
Then we are back to the expression for 
$\int\d^3x\,{\cal L}_B$ used in the main text; cf. (21, 22) of the main text. 

\subsection{2. Electrostatic energy for point charges}

\comment{ 
We now calculate the electrostatic energy change
$-\frac{1}{2}\frac{{\rm d}}{{\rm d} t}\int{\rm d}^3\,x\,\rho\di\rho$ for
$N$ point-like charges with coordinates $r_\alpha(t)$ and charge
$e_\alpha$. One first smears their density:
$\rho(x,t)=\sum_{\alpha=1}^Ne_\alpha\hat\delta(x-r_\alpha(t))$, where
$\hat\delta(x)$ is an approximate $\delta$-function with a finite
$\hat\delta(0)$. Then (\ref{aba2}) implies \begin{align} &-\frac{{\rm
d}}{{\rm d} t}\int{\rm d}^3\,x\,\rho\di\rho =\sum_{\alpha\not=\beta}
\frac{{\rm d}}{{\rm d} t}\int\frac{{\rm d}^3\,x\,{\rm
d}^3\,y}{4\pi}\,\frac{e_\alpha e_\beta}{|x-y|}\times\nonumber\\
&\hat\delta(x-r_\alpha(t))\,\hat\delta(x-r_\beta(t))=\sum_{\alpha\not=\beta}\frac{{\rm
d}}{{\rm d} t} \frac{e_\alpha e_\beta}{|r_\alpha(t)-r_\beta(t)|}.
\end{align} where $\hat\delta(x)\to\delta(x)$ is taken only in the last
equality.  }

Here we recall how to calculate the change of the electrostatic energy:
\BEA
\label{s1}
\frac{\d E_S}{\d t}
&=&-\frac{1}{2}\frac{{\rm d}}{{\rm d} t}\int{\rm d}^3\,x\,\rho\di\rho\\
&=&\frac{1}{2}\frac{{\rm d}}{{\rm d} t}\int\frac{ {\rm d}^3\,x\, {\rm d}^3\,x}{4\pi}
\, \frac{\rho(x,t)\rho(y,t)}{|x-y|},
\label{s11}
\EEA
for point charges. The point here is that for point charges (\ref{s1}) is 
infinite, but its change in time is finite, since infnities cancel out. 

The charge density for $N$ points with charges $e_\a$ and coordinate vectors $r_\a$
reads
\BEA
\label{s2}
\rho(x,t)=\sum_{\alpha=1}^Ne_\alpha\delta(x-r_\alpha(t)).
\EEA
We shall temporarily move from (\ref{s2}) to regularized $\delta$-functions 
$\hat\delta(x-r_\alpha(t))$, where $\hat\delta(0)$ is finite:
\BEA
\label{s3}
\rho(x,t)=\sum_{\alpha=1}^Ne_\alpha\hat\delta(x-r_\alpha(t)).
\EEA
Putting (\ref{s3}) into (\ref{s11}) we get
\begin{gather}
\label{s4}
\frac{\d E_S}{\d t}=\sum_{\a}
\frac{1}{2}\frac{{\rm d}}{{\rm d} t}\int\frac{ {\rm d}^3\,x\, {\rm d}^3\,x}{4\pi}
\, \frac{e^2_\a \hat\delta(x-r_\a(t))\hat\delta(y-r_\a(t))}{|x-y|}\\
\label{s5}
+\sum_{\a\not=\b}
\frac{1}{2}
\frac{{\rm d}}{{\rm d} t}\int\frac{ {\rm d}^3\,x\, {\rm d}^3\,x}{4\pi}
\, \frac{e_\a e_\b\hat\delta(x-r_\a(t))\hat\delta(y-r_\b(t))}{|x-y|}.
\end{gather}
It is seen that (\ref{s4}) nullifies, since the integral is finite and does not depend on time (via $r_\a(t)$). 
In (\ref{s5}) we can take the regularization out:
\BEA
\label{s6}
\frac{\d E_S}{\d t}=\frac{1}{2}\,\frac{\d }{\d t}\sum_{\a\not=\b}\frac{e_\a e_\b}{|r_\a(t)-r_\b(t)|},
\EEA
which is the final and finite result. 

\subsection{3. Massive electrodynamics}

\subsubsection{3.1 Equations of motion and Hamiltonian}

The Proca Lagrangian $\int{\rm d}^3\,x\, {\cal L}$ of 
electrodynamics with mass $m$ reads \cite{mass,stue}
\begin{gather}
{\cal L}_m=\frac{1}{2}(\partial_i \phi+\dot A_i)^2-\frac{1}{2}B_i^2
+\frac{m^2}{2}(\phi^2-A_i^2)-\rho\phi+J_iA_i.
\label{lagrom}
\end{gather}
This theory is not gauge-invariant. Equations of motion read
\bea
\label{agnus10}
&&\partial_k \dot\phi  +\ddot A_k=\Delta A_k-\pa_k (\pa_i A_i)+J_k-m^2A_k,\\
&&\Delta\phi-m^2\phi=-\rho-\pa_i\dot A_i.
\label{agnus20}
\eea
Apply $\pa_t$ to (\ref{agnus20}) and $\pa_k$ to (\ref{agnus10}). Together these lead to
\bea
\label{rev}
\dot\rho  +\pa_kJ_k=m^2(\dot\phi+\pa_kA_k).
\eea
Hence if we impose the charge conservation $\dot\rho  +\pa_kJ_k=0$, then (\ref{rev}) together with $m>0$ leads to 
the Lorenz gauge \cite{mass,stue}
\bea
\label{loro}
\dot\phi+\pa_kA_k=0.
\eea
Using (\ref{loro}) we present (\ref{agnus20}) and (\ref{agnus10}) as
\bea
\label{uro}
&& \Delta\phi-m^2\phi=-\rho+\ddot{\phi},\\
&& \Delta A_k-m^2 A_k=-J_k+\ddot{A}_k,
\label{kuro}
\eea
i.e. $\phi$ and $A_k$ are decoupled up to (\ref{loro}).
 
Let us turn to Hamiltonizing (\ref{lagrom}). We write 
\bea
\label{vem}
{\cal H}_m=p_i\dot{A}_i - {\cal L}_m,
\eea
and via the Legendre transform:
\bea
\label{gamelo}
\frac{\delta}{\delta \dot A_k(y)}\int{\rm d}^3\,x\, {\cal H}_m=0,\quad
\frac{\delta}{\delta \phi(y)}\int{\rm d}^3\,x\, {\cal H}_m=0,
\label{aleppo}
\eea
exclude four variables $\dot{A}_i$ and $\phi$ in favor of three momenta $p_i$ \cite{gitman}:
\bea
\label{blum}
&& \phi=\frac{1}{m^2}(\rho+\pa_ip_i), \\
&& p_i=\dot A_i +\frac{1}{m^2}\pa_k(\rho+\pa_ip_i).
\label{blume}
\eea
Putting (\ref{blum}, \ref{blume}) into (\ref{vem}) we get:
\begin{gather}
\label{veno}
{\cal H}_m=\frac{1}{2}p_i^2+\frac{1}{2}B_i^2+\frac{m^2}{2}A_i^2+\frac{1}{2m^2}(\rho+\pa_ip_i)^2-J_iA_i,
\end{gather}
where $p_i$ and $A_i$ are (resp.) independent canonic momenta and coordinates:
\bea
\label{la1}
\dot p_j=\Delta A_j-\pa_j \pa_kA_k-m^2A_j+J_j,\\
\dot{A}_j=p_j-\frac{1}{m^2}\pa_j (\rho+\pa_ip_i).
\label{la2}
\eea
For the time-derivative we get
\bea
\frac{{\rm d}}{{\rm d} t}\int{\rm d}^3 x\, {\cal H}_m&=&\int{\rm d}^3 x\,[\, \frac{\dot\rho}{m^2}(\rho+\pa_ip_i)-\dot J_iA_i\,],\nonumber\\
&=&\int{\rm d}^3 x\,[\, \dot\rho\phi-\dot J_iA_i\,],
\label{vera}
\eea
where we employed (\ref{blume}). Now (\ref{vera}) is finite and 
well-defined for $m\to 0$ [cf.~(\ref{uro}, \ref{kuro})],
where it amounts to using the Lorenz gauge.

\subsubsection{3.2 Exclusion of $\phi$}

By analogy to the massless situation we can attempt to exclude $\phi$ in the Lagrangian via (\ref{agnus20}):
\bea
\label{babo}
&& \phi=-\Gamma^{-1}(\rho+\pa_i \dot A_i), \\ 
&& \Gamma\equiv \Delta-m^2, \\ 
&& (\Gamma^{-1}f)({x})=-\int\frac{{\rm d}^3 y}{4\pi} \,\frac{f(y)}{|{x}-{y}|}\,e^{-m| x- y|}.
\eea
To this end we integrate (\ref{lagrom}) by parts and write it as
\bea
\label{lagromm}
\int\d^3x\,{\cal L}_m=\int\d^3x\,[\, -\frac{1}{2}\phi\Gamma\phi -\rho\phi -\phi\pa_i\dot A_i \\
+\frac{1}{2}\dot A_i^2 -\frac{1}{2}B_i^2-\frac{m^2}{2}A_i^2+J_iA_i\,].
\eea
Now the first three terms in the RHS of (\ref{lagromm}) are to be transformed via (\ref{babo}),
and 
\bea
\label{oli}
\int{\rm d}^3 x\,a(x) (\Gamma^{-1}b)(x) = \int{\rm d}^3 x\,b(x) (\Gamma^{-1}a)(x).
\eea
We get
\BEA
\label{mmm}
&\int\d^3x\,{\cal L}_m=\int\d^3x\,[\,
\frac{1}{2}(\pa_i\dot A_i)\Gamma^{-1} (\pa_k\dot A_k)+ (\pa_i\dot A_i)\Gamma^{-1}(\rho) \nonumber\\
&-\frac{1}{2}\rho\Gamma^{-1}\rho +\frac{1}{2}\dot A_i^2 -\frac{1}{2}B_i^2-\frac{m^2}{2}A_i^2+J_iA_i\,].
\EEA
Now we transform the term $(\pa_i\dot A_i)\Gamma^{-1}(\rho)$, since eventually we aim at reproducing the autonomous dynamics of $A_i$
given by (\ref{kuro}) that does not contain $\rho$. We neglect the full time-derivative, employ the charge conservation $\dot \rho+\pa_k J_k=0$
and replace $(\pa_i\dot A_i)\Gamma^{-1}(\rho)$ by $(\pa_i A_i)\Gamma^{-1}(\pa_k J_k)$. Hence the transformed Lagrangian reads:
\begin{gather}
\int\d^3x\,{\cal L}'_m =\int\d^3x\,[\,-\frac{1}{2}\rho\Gamma^{-1}\rho +\frac{1}{2}(\pa_i\dot A_i)\Gamma^{-1} (\pa_k\dot A_k) \nonumber\\
+ (\pa_i A_i)\Gamma^{-1}(\pa_k J_k)+\frac{1}{2}\dot A_i^2 
-\frac{1}{2}B_i^2-\frac{m^2}{2}A_i^2+J_iA_i].
\label{bingo}
\end{gather}
Note a relation 
\BEA
\label{kot}
&-\int\d^3x\,\dot B_i \Gamma^{-1}\dot B_i \nonumber\\
&=\int\d^3x\,[\, (\pa_k\dot A_k) \Gamma^{-1}(\pa_i\dot A_i)+\dot A_k^2 +m^2 \dot A_k \Gamma^{-1}\dot A_k\,],
\EEA
whose analogue was employed by us for the $m=0$ situation. 
Now employing (\ref{kot}) is not going to be useful, since the autonomous description for 
the $m>0$ case is provided directly by $A_i$ and not by $B_i$.

At any rate (\ref{bingo}) contains only $A_i$, $\dot A_i$ and $J_i$ and we can look for Lagrange equations generated by it treating 
$A_i$, $\dot A_i$ and $J_i$ as (resp.) coordinates, velocities and external fields [cf.~(\ref{agnus10})]:
\bea
\label{bolivar}
&\ddot A_i-\Delta A_i+m^2 A_i-J_i \nonumber\\
&= \Gamma^{-1} (\pa_i \pa_k\ddot A_k)-\Gamma^{-1} (\pa_i \pa_kJ_k)-\pa_i\pa_kA_k.
\eea
This is an integro-differential equation. Its LHS compares with
(\ref{kuro}), but its RHS is generally not zero. We conclude that
(\ref{bingo}) does not correspond to the autonomous description of $A_i$
that is given by (\ref{kuro}). While such a Lagrangian can be written
down on the basis of (\ref{kuro}), it does not relate to the original
Lagrangian (\ref{lagrom}). 

\subsection{4. Natural configurations of charge density and current}

As for any vector, one can apply the Helmholtz's theorem (obtained e.g.
via the Fourier representation) for representing the current $J_k$ as
\BEA
J_i=J_i^\perp+J_i^\parallel, \\
\pa_iJ_i^\perp=\epsilon_{ijk}\pa_jJ_k^\parallel=0.
\label{rashid}
\EEA
Let us now assume that $J_i^\parallel$ does not depend on time: $\dot J_i^\parallel=0$.
Using the continuity equation $\dot \rho+\pa_kJ_k=\dot \rho+\pa_kJ^\parallel_k=0$ and (\ref{rashid}), we obtain 
\BEA
\label{swo}
\rho(x,t)=\rho(x,0)- t\, \pa_kJ_k^\parallel (x).
\EEA
If we demand that all involved charge densities stay bounded for any time $t$, then (\ref{swo}) leads to $\dot\rho=0$, i.e. to
$\pa_kJ_k^\parallel (x)=0$ and hence to $J_k^\parallel (x)=0$. 

Thus, under a natural additional condition, we conclude that stationary currents are vortical:
$\dot J_i^\parallel=0$ leads to $J_k^\parallel (x)=0$. 

\comment{Let us argue that $\dot J_i=0$ leads to $\dot H_B=0$. For any
vector $J_i$ the Helmholtz's theorem (obtained e.g. via the Fourier
representation) implies $J_i=J_i^\perp+J_i^\parallel$, where
$\pa_iJ_i^\perp=\epsilon_{ijk}\pa_jJ_k^\parallel=0$. Now (\ref{cons})
shows that $\dot J_k=0=\dot J_k^\parallel$ leads to $\dot\rho=0$.
Otherwise, $\rho$ will grow as $\rho\propto t$, which is unphysical
\cite{ribaric}. In its turn $\dot\rho(x,t)=0$ leads to two conclusions.
First, together with $\dot J_i=0$ we get $\dot H_B=0$ from
(\ref{dan}). Second, we obtain $J_k^\parallel(x,t)=0$ explaining why the
conserving (\ref{golem}) contains only $J^\perp_k$ via
$R_i=\epsilon_{ijk}\pa_jJ^\perp_k$. 
}

\subsection{5. Calculation of $\dot H_D$ in the Lorenz gauge}

\subsubsection{5.1 The Lorenz gauge}
 
We return to the rate of the standard EMF Hamiltonian 
\BEA
\dot H_{D}=\int{\rm d}^3\,x\,\left[\phi\dot\rho-\dot J_iA_i\right],
\label{ser2}
\EEA
which is gauge-variant. We shall calculate (\ref{ser2}) in the Lorenz gauge 
\BEA
\dot\phi+\pa_kA_k=0, 
\label{lorenz}
\EEA
and for slow and space-localized sources $\rho$ and $J_k$. 
The Lorenz gauge is selected, because it emerges in the limit $m\to 0$ 
of the massive electrodynamics and because it is relativistically covariant. 
The purpose of the calculation is to check whether our conclusions on
the second law can be seen on the level of (\ref{ser2}, \ref{lorenz}). 

Here are the known retarded-potential solutions that hold (\ref{lorenz}):
\bea
\label{reto2}
\phi({x}, t)=\frac{1}{4\pi}\int\frac{{\rm d}^3\,y}{|{x}-{y}|} \, 
\rho({y}, t-|{x}-{y}|), \\
A_i({x}, t)=\frac{1}{4\pi}\int\frac{{\rm d}^3\,y}{|{x}-{y}|} \, 
J_i({y}, t-|{x}-{y}|).·
\label{reto3}
\eea

\subsubsection{5.2 The space-localized and slow (non-relativistic) approximation.}

The space-localized approximation amounts to the following points
elucidated on the example of (\ref{reto3}). First, $J_i({{y}, t})$ as a
function of ${y}$ is well-localized in the vicinity of (say) ${y}=0$.
Hence the main contribution to the $\int{\rm d}^3\,x\,$ integral in
(\ref{reto3}) comes from $|{x}|\gg 1$. We can put in the RHS of
(\ref{reto3}): ${|{x}-{z}|}\approx {|{x}|}$ and ${|{x}-{y}|}\approx
{|{x}|}$. The second assumption is that $J_i({{y}, t})$ is a slow
function of $t$, hence $J_i({y}, t-|{x}-{y}|)\approx J_i({y}, t-|{x}|)$.
Hence
\bea
\label{reto4}
\phi({x}, t)&\simeq&\frac{1}{4\pi|x|}\int{{\rm d}^3\,y}\,
\rho({y}, t-|{x}|)\\
\label{reto44}
&=&\frac{1}{4\pi|x|}\int{{\rm d}^3\,y}\,
\rho({y}, t), \\
A_i({x}, t)&\simeq&\frac{1}{4\pi |x|}\int{{\rm d}^3\,y}\,
J_i({y}, t-|{x}|)·
\label{reto5}
\eea
where in (\ref{reto44}) we addtionally used the
charge conservation. It is now seen that within this approximation we
can put
\begin{gather}
\int{\rm d}^3\,x\,\phi\dot\rho
\simeq \int{\rm d}^3\,x\,\frac{\dot\rho(x,t)}{4\pi |x|}\int{\rm d}^3\,y\,\rho(y,t)\nonumber\\
\simeq \int{\rm d}^3\,x\,{\rm d}^3\,y\, \frac{\dot\rho(x,t)\rho(y,t) }{4\pi |x-y|}
=-\frac{1}{2}\frac{\d}{\d t} \int{\rm d}^3\,x\,\rho\di \rho.
\end{gather}
Hence the factor $\phi\dot\rho$ in (\ref{ser2}) approximately recovers the electrostatic energy change 
seen also for $\dot H_B$; see (26, 27) of the main text. 

We now turn to (\ref{reto5}) and the term $A_i\dot J_i$ in (\ref{ser2}):
\begin{gather}
\label{mango}
\int{\rm d}^3\,x\,\dot J_iA_i =\int_0^\infty\frac{u^2\d u}{4\pi}\int_0^t r\,\d r\, \dot f_i(r,t)\,f_i(u,t-r),\\
f_i(r,t)\equiv\int\d \Omega \,J_i(r,\Omega,t),
\end{gather}
where we went to spherical variables, $\int\d \Omega$ is the integration
over the spherical angles, and where we assumed 
\BEA
\label{aq1}
J_i(x,t)=0\qquad {\rm for}\qquad t<0.
\EEA
We now check the sign of (\ref{mango}) under an additional condition
\BEA
\label{aq2}
J_i(x,\tau)=0. 
\EEA
Eqs.~(\ref{aq1}, \ref{aq2}) amount to a cyclic change. 
Now (\ref{mango}) does not have a definite sign, as can be
seen e.g. by taking $f_i(r,t)=\delta_{i1}a(r)b(t)$, where $a(r)$ is a function 
well-localized at $r\simeq r_0$, and where $b(t)=\sin t$, $\tau=2\pi$.
Then (\ref{mango}) changes its sign as a function of $r_0$ for $r_0<2\pi$.


\begin{thebibliography}{99}
\bibitem{comment}From the huge literature on electromagnetism we
selected references that stress its evolving, open-ended nature.
\bibitem{comment2}
Neglecting back-reaction does not mean neglecting the energy
transfer which is the very point of defining work \cite{balian}.  Taking
the fullback-reaction into account leads to exceedingly complex theories.
Moreover, for electrodynamics such a theory would be not completely
consistent \cite{landau}.
\bibitem{lagron}Two Lagrangians that differ by a full time-derivative are
equivalent \cite{landau}. But the inverse does not hold: two Lagrangians
(also for EMF \cite{sudbery}) can generate the same equations of motion
without differing from each other by a full time-derivative \cite{sudbery,okubo}.
\end{thebibliography}
\end{document}